\documentclass[iop]{emulateapj}

\usepackage{graphicx}
\usepackage{txfonts}
\usepackage{natbib}
\usepackage{amssymb}
\usepackage{textcomp}
\usepackage{float}
\usepackage{gensymb}
\DeclareRobustCommand{\ion}[2]{%
\relax\ifmmode
\ifx\testbx\f@series
{\mathbf{#1\,\mathsc{#2}}}\else
{\mathrm{#1\,\mathsc{#2}}}\fi
\else\textup{#1\,{\mdseries\textsc{#2}}}%
\fi}

\begin{document}

\title{Spatially-Resolved Star Formation Main Sequence Of Galaxies in the CALIFA Survey}

\author{M. Cano-D\'{\i}az$^{1}$, S.F. S\'anchez$^{1}$, S. Zibetti$^{2}$, Y. Ascasibar$^{3,4}$, J. Bland-Hawthorn$^{5}$, B. Ziegler$^{6}$, R. M. Gonz\'alez Delgado$^{7}$, C.J. Walcher$^{8}$, R. Garc\'{\i}a-Benito$^{7}$, D. Mast$^{9,10}$, M.A. Mendoza-P\'erez$^{7}$, J. Falc\'on-Barroso$^{11,12}$, L. Galbany$^{13,14}$, B. Husemann$^{15}$, C. Kehrig$^{7}$, R. A. Marino$^{16,17}$, P. S\'anchez-Bl\'azquez$^{3,18}$, C. L\'opez-Cob\'a$^{1}$, \'A. R. L\'{o}pez-S\'{a}nchez$^{19,20}$, J.M. Vilchez$^{7}$ \\
$^{1}$ Instituto de Astronom\'{\i}a, Universidad Nacional Aut\'onoma de M\'exico, Apartado Postal 70-264, Mexico D.F., 04510 Mexico \\
$^{2}$ INAF-Osservatorio Astrofisico di Arcetri, Largo Enrico Fermi 5, 50125 Firenze, Italy \\
$^{3}$ Departamento de F\'{\i}sica Te\'orica, Universidad Aut\'onoma de Madrid, E-28049 Madrid, Spain \\
$^{4}$ Astro-UAM, UAM, Unidad Asociada CSIC \\
$^{5}$ Sydney Institute for Astronomy, School of Physics, University of Sydney, NSW 2006, Australia \\
$^{6}$ Universit\"{at} Wien, Institut f\"{u}r Astrophysik, T\"{u}rkenschanzstra$\ss$e 17, A-1180 Wien, Austria \\
$^{7}$ Instituto de Astrof\'{\i}sica de Andaluc\'{\i}a (IAA-CSIC), Glorieta de la Astronom\'{\i}a s/n Aptdo. 3004, 18008 Granada, Spain \\
$^{8}$ Leibniz-Institut f\"ur Astrophysik Potsdam (AIP), An der Sternwarte 16, D-14482 Potsdam, Germany \\
$^{9}$ Observatorio Astron\'omico, Laprida 854, X5000BGR, C\'ordoba, Argentina \\
$^{10}$ Consejo de Investigaciones Cient\'{\i}ficas y T\'ecnicas de la Rep\'ublica Argentina, Avda. Rivadavia 1917, C1033AAJ, CABA, Argentina \\
$^{11}$ Instituto de Astrof\'{\i}sica de Canarias, V\'{\i}a L\'actea s/n, 38205 La Laguna, Tenerife, Spain \\
$^{12}$ Departamento de Astrof\'{\i}sica, Universidad de La Laguna, 38205 La Laguna, Tenerife, Spain \\
$^{13}$Millennium Institute of Astrophysics MAS, Nuncio Monse\~{n}or S\'otero Sanz 100, Providencia, 7500011 Santiago, Chile\\
$^{14}$Departamento de Astronom\'{\i}a, Universidad de Chile, Camino El Observatorio 1515, Las Condes, Santiago, Chile\\
$^{15}$European Southern Observatory, Karl-Schwarzschild-Str. 2, 85748 Garching b. M\"unchen, Germany\\
$^{16}$Dep. de Astrof\'{\i}sica y CC. de la Atm\'osfera, Facultad de CC. F\'{\i}sicas, UCM, Avda. Complutense s/n, 28040 Madrid, Spain\\
$^{17}$Department of Physics, Institute for Astronomy, ETH Z\"{u}rich, CH-8093 Z\"{u}rich, Switzerland \\
$^{18}$Instituto de Astrofis\'{\i}ca, Facultad de F\'{\i}sica, Pontificia Universidad Cat\'olica de Chile, Casilla 306, Santiago 22, Chile\\
$^{19}$Australian Astronomical Observatory, PO Box 915, North Ryde, NSW 1670,Australia\\
$^{20}$Department of Physics and Astronomy, Macquarie University, NSW 2109, Australia\\
}


\begin{abstract}
The ``main sequence of galaxies'' $-$ defined in terms of the total star formation rate $\psi$ 
vs. the total stellar mass $M_*$ $-$ is a well-studied tight relation that has been observed at several 
wavelengths and at different redshifts. All earlier studies have derived this relation from integrated properties of galaxies. We recover the same relation from an analysis of spatially-resolved properties, with integral field spectroscopic (IFS) observations of 306 galaxies from the CALIFA survey. We consider the SFR surface density in units of log(M$_{\odot}$ yr$^{-1}$ Kpc$^{-2}$) and the stellar mass surface density in units of log(M$_{\odot}$ Kpc$^{-2}$) in individual spaxels which probe spatial scales of 0.5-1.5 Kpc. This local relation exhibits a high degree of correlation with small scatter ($\sigma = 0.23$ dex), irrespective of the dominant ionisation source of the host galaxy or its integrated stellar mass. We highlight: $(i)$ the integrated star formation main sequence formed by galaxies whose dominant ionisation process is related to star formation, for which we find a slope of 0.81 $\pm 0.02$; (ii) the spatially-resolved relation obtained with the spaxel analysis, we find a slope of 0.72 $\pm 0.04$; (iii) for the integrated main sequence we identified also a sequence formed by galaxies that are dominated by an old stellar population, which we have called the retired galaxies sequence.
\end{abstract}


\section{Introduction} \label{Intro}

Thanks to the increasing number of statistical studies of both local and distant galaxies, it has been possible to reveal and confirm several correlations in extragalactic astronomy. One of these relations is the so-called star formation main sequence (SFMS) of actively star forming galaxies, which relates the star formation rate (SFR, $\psi$) and the stellar mass ($M_{*}$).

The SFMS is an approximately linear correlation between log$\psi$ and log$M_{*}$, that has been observed in the local universe as well as at high redshifts. See for example:  \citet{Brinchmann04}, \citet{Salim07}, \citet{Renzini15} (hearafter RP15) for $z \sim 0$, \citet{Peng10} for $z\lesssim1$ and \citet{Daddi07} (hearafter D07) for $z>1$. In particular, \citet{Speagle14} (hearafter S14) performed a compilation of SFMS relations reported in the literature and shows a summary of the evolutionary behaviour of the SFMS with redshift (up to $z\sim$ $6$). \citet{Katsianis15} performed a similar study of the evolution of this correlation for $z\sim 1-4$. The SFMS is also recovered in cosmological simulations \citep[hearafter S15]{Dave11,Torrey14, Sparre15}. In fact, this correlation has been proven to be tight, with a scatter of $\sim0.2-0.35$ dex in observations and in theoretical predictions (D07, S15, S14). The values of the slope and zero points for the SFMS may vary within a large range in the literature (see S14 for a compilation). It has been proposed that this variations may originate from the selection criteria used to select star forming galaxies. RP15 has proposed a new objective definition for the SFMS for local galaxies, which favours values of 0.76 dex and -7.64 log(M$_{\odot}$ yr$^{-1}$) for the slope and zero points respectively.

Although the physical distinction between star-forming and passive galaxies is not straightforward \citep[see e.g.][]{Casado15}, the SFMS correlation provides a convenient way to classify galaxies in terms of their SF properties. Generally, the star-forming galaxies lie on the main sequence, red elliptical galaxies tend to lie below the relation, while the starburst galaxies, lie above the sequence (S15).

All previous SFMS studies have been done, using integrated quantities for the galaxies. Due to observational restrictions, the derivation of quantities like SFR and M$_{*}$ have been performed using, for example, single fiber spectroscopy affected by aperture losses that need to be corrected. Integral Field Spectroscopy (IFS) allow to have spatial and spectroscopical information for extended objects, as it delivers individual spectra for each point of the observed target (restricted to the instrument spatial resolution), which makes IFS a suitable observational technique to study of spatially-resolved physical quantities in galaxies. Studying the SFMS with IFS would allow to obtain both, integrated and local correlations. If the field of view covers the entire optical extension of the galaxy, these data are not affected by aperture losses.

IFS when used to perform sky surveys, allows to do statistical studies of spatially-resolved physical quantities of galaxies. The Calar Alto Legacy Integral Field Area survey CALIFA, \citep{Sanchez12} is an ongoing extragalactic optical IFS survey, designed to observe around $600$ galaxies, which makes it suitable to perform spatially-resolved studies with statistical significance.

We present the results of studying the spatially-resolved SFMS based on data from the CALIFA survey. Throughout this paper we have adopted a Salpeter IMF and a cosmology defined by: H$_{0}$= 71 km s$^{-1}$ Mpc$^{-1}$, $\Omega_{\Lambda}$=0.7 and a flat Universe.


\section{Data and Sample of galaxies} \label{Data}

We used the available sample observed by \mbox{CALIFA} until February 2015, consisting of 535 galaxies that are representative of its mother sample \citep[][mass and redshift ranges are $10^{9.7} \textless M_{*} \textless 10^{11.4}$ M$_{\odot}$ and $0.005 \textless$ $z$ $\textless 0.03$ respectively]{Walcher14}, which includes galaxies of all morphological types, inclinations and environments. Galaxies from CALIFA extended sample observed as ancillary programs, i.e. galaxies not contemplated in the original mother sample, were also included. For this reason the $M_{*}$ of some of the galaxies in our sample may be lower than the limits established on the mother sample. In order to avoid inclination effects, we clipped our sample to a low inclination ($i$) galaxies subsample with $i<$ 60$\textdegree$, for which 306 galaxies remained.

The observations were performed using the PMAS instrument \citep{Roth05} in the PPAK configuration \citep{Kelz06}. The observing strategy guarantees a complete coverage of the spatial extension of the galaxies up to 2.5 effective radius, with a FWHM$\sim2.5''$ \citep{Garcia15}, which corresponds to $\sim1$ kpc at the average redshift of the survey \citep[for further information of the survey, sample and observational strategy see:][]{Sanchez12}

We used data from the V500 setup that covers a wavelength range of $3745$ to $7300$ $\AA$, with a nominal resolution of  $\lambda\textfractionsolidus\Delta\lambda$= $850$ at $4500$ $\AA$. The datacubes were provided by version 1.5 of the pipeline \citep{Garcia15}, and consist of a regular grid of ~72$\times$78 spectra, with a 1"/spaxel size centred in the galaxies.

\section{Star Formation Rates and Stellar Mass Calculation} \label{SFR}

The datacubes were analysed using the Pipe3D pipeline \citep{Sanchez16}, which is a tool that fits the continuum with stellar population models and measures the nebular emission lines. This pipeline is based on the Fit3D fitting tool \citep{Sanchez15}. For this particular implementation we adopted the GSD156 library of simple stellar population models \citep[hereafter CF13]{CidFernandes13}, that comprises 156 templates covering 39 stellar ages (from 1Myr to 13Gyr), and 4 metallicities ($Z/Z_{\odot}=$ 0.2, 0.4, 1, and 1.5). This templates have been extensively used within the CALIFA collaboration \citep[e.g.][CF13]{Perez13, GonzalezDelgado14}. Details of the fitting procedure, dust attenuation curve, and uncertainties of the process are given in \citep{Sanchez15,Sanchez16}. 

We applied a spatial binning to each datacube to reach an homogenous S/N of 50 across the field of view. Then, the stellar population fitting was applied to the coadded spectra within each spatial bin. Finally, following the procedures described in CF13 and S\'anchez et al. 2015b, we derive the stellar-population model for each spaxel by re-scaling the best fitted model within each spatial bin to the continuum flux intensity in the corresponding spaxel. The stellar-population model is then substracted to create a gas-pure cube comprising only the ionised gas emission lines. Thus it is assumed the same M/L ratio and dust attenuation for those spaxels within the same spatial-bin, to derive a spaxel-wise map of any stellar property, like the $M_{*}$ surface density ($\Sigma_{*}$).

For the gas-pure cube, the strongest emission lines within the considered wavelength range, including H$\alpha$ and H$\beta$ are fitted spaxel-by-spaxel to derive their corresponding flux intensity and equivalent width (EW) maps. The H$\alpha$ flux is then corrected by the ionised gas dust attenuation, derived using the Balmer decrement assuming a canonical value of 2.86, adopting a  \citet{Cardelli89} extinction law and $R_{v}=3.1$. The H$\alpha$ luminosity distribution is derived by correcting for the cosmological distance and, by applying the \citet{Kennicutt98} conversion from H$\alpha$ luminosity to SFR we derive both the integrated SFRs and the spaxel-wise distribution of the SFR surface density ($\mu$SFR). The measured $\mu$SFR and $\Sigma_{*}$ have not been corrected for galaxy inclination. However, since we have limited our sample to galaxies with $i<$ 60$\textdegree$, the impact of inclination on the projected surface densities will be less than a factor of two.

\section{Results} \label{Results}

The procedure described in Section \ref{SFR} was applied to the 306 galaxies mentioned in Section \ref{Data} to construct the SFMS relation for the CALIFA sample.

\subsection{Integrated SFMS relation}

\begin{figure}
  \centering
    \includegraphics[width=0.35\textwidth, angle=90]{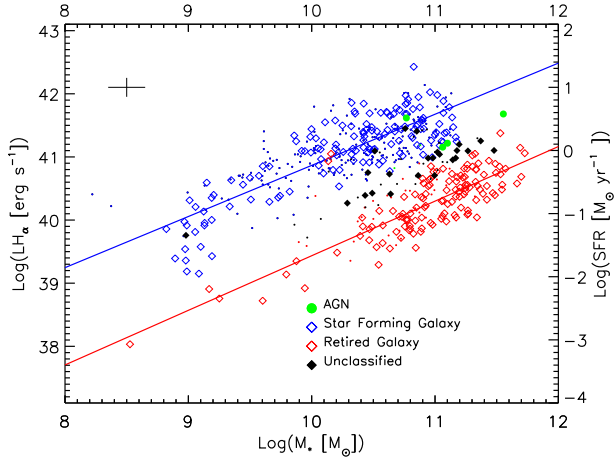}
  \caption{Integrated $M_{*}$vs.SFR relation for the CALIFA sample. Green symbols represent galaxies that lie above the Kewley Limit (KL) and whose H$\alpha$ Equivalent Widths (EW) are $>$ 6$\AA$, i.e. are galaxies whose ionisation emission is dominated by the AGN activity. Blue symbols represent galaxies below the KL and with EW(H$\alpha$) $>$ 6$\AA$, and that have inclinations $<$ 60$\textdegree$, i.e. that lie in the Star Formation Main Sequence (SFMS). Red symbols represent galaxies with EW(H$\alpha$) $<$ 3$\AA$, i.e. that lie in a Retired Galaxies Sequence (RGS). Finally black symbols account for galaxies whose EW(H$\alpha$) lie between 3$\AA$ and 6$\AA$, i.e. galaxies whose dominant ionisation process is uncertain. Shaped symbols hold for galaxies in the low inclination subsample, that are the ones used for the rest of the analysis, however for completeness, the points show the complementary high inclination subsample. Blue and red lines are the linear fittings to the SFMS for low inclined galaxies and RGS respectively (See Table \ref{table1} for details). SFRs shown in the plot only hold for the SF dominated galaxies, for the rest these rates are only the result of the linear transformation of the H$\alpha$ luminosities with the \citet{Kennicutt98} relation. For further details of the galaxies classification please refer to Section 4.1.}
  \label{fig:sfrVSmas_int}
\end{figure}

Figure \ref{fig:sfrVSmas_int}, shows the integrated $M_{*}$ vs. SFR relation that was constructed integrating the corresponding spatially-resolved quantities of all the galaxies in the sample. Preliminary versions of this plot for smaller subsamples of the CALIFA data have been already presented in \citet{Sanchez13} and \citet{Catalan2015}. 

In Figure \ref{fig:sfrVSmas_int} we classify the dominant ionisation source in the galaxies based on a combination of EWs(H$\alpha$) classification, introduced by \citet{CidFernandes11} and a classification using the Kewley demarcation limit (KL) \citep{Kewley01} in the Baldwin-Phillips-Terlevich (BPT) diagram \citep{Baldwin81}: (i) the green symbols account for galaxies for which the [\ion{O}{iii}]/H$\beta$ and  [\ion{N}{ii}]/H$\alpha$ line ratios of their integrated ionised gas emission lines lie above the KL and whose EW(H$\alpha$) are $\textgreater$ 6$\AA$, meaning that the dominant ionisation process in these galaxies comes from the nuclear activity. (ii) The blue symbols represent the galaxies that lie below the KL in the BPT diagram, and whose  EW(H$\alpha$) are $\textgreater$ 6$\AA$, which means that the ionisation processes in these galaxies are dominated by SF, as it is shown in Figure 3 of \citet{Sanchez14}. (iii) The red symbols account for galaxies whose EW(H$\alpha$) are $\textless$ 3$\AA$ regardless of their position in the BPT diagram, meaning that most probably the dominant processes in the gas ionisation come from an old stellar population \citep{CidFernandes11}. (iv) Finally, black symbols represent galaxies whose EW(H$\alpha$) are:  3$\AA$ $\textless$ EW $\textless$ 6$\AA$, which means that their main ionisation process remain uncertain. Shaped symbols account for the galaxies from the low $i$ subsample, that are the ones used for the rest of the analysis. For completeness, points show the galaxies for the complementary high $i$ subsample.

Although, we have transformed the H$\alpha$ luminosities to SFRs for all the galaxies adopting the  \citet{Kennicutt98} relation, we should clarify that the interpretation of SFRs is only valid for galaxies (and regions) that are ionised by young stars (following our previous criteria). For the rest of the galaxies, the SFRs presented in the plot are indeed just a linear transformation of the H$\alpha$ luminosity.  

For the star-forming and retired galaxies sequences (blue and red respectively), which, from now on will be referred to as SFMS and Retired Galaxies Sequence (RGS), we have fitted linear correlations between log$(\psi)$ and log$(M_{*})$ whose characteristics are listed in Table \ref{table1}, along with their dispersions. The slope and zero point for the SFMS are in good agreement with the values reported in the literature for local galaxies (RP15, S14 and references therein. For further details see section 5). For the black and green points no further analysis was performed due to the uncertain main ionisation process of the gas and due to its poor statistics, respectively.

\begin{table*}
 \begin{center}
    \begin{tabular}{ l  c  c  c  c p{5cm}}
    \hline
    \hline
    Coefficient & SFMS & RGS & \shortstack{Spatially-Resolved SFMS \\ 100$\%$ of the data} & \shortstack{Spatially-Resolved SFMS \\ 80$\%$ of the data}\\ \hline
    Pearson Correlation Coeff. ($\rho$) & 0.84 & 0.85 & 0.61 & 0.63 \\ \hline
    $\rho$ 99$\%$ Confidence Interval & (0.76, 0.89) & (0.77, 0.90) & (0.60, 0.61) & (0.62, 0.63) \\ \hline
    Slope & 0.81$\pm 0.02$ & 0.86$\pm 0.02$ & 0.68$\pm0.04$ & 0.72$\pm0.04$ \\ \hline
    Zero Point & -8.34$\pm 0.19^{\textasteriskcentered}$ & -10.32$\pm 0.24^{\textasteriskcentered}$  & -7.63$\pm0.34^{\dag}$ & -7.95$\pm0.29^{\dag}$ \\ \hline
    Standard Deviation ($\sigma$) & 0.20 & 0.22  & 0.23 & 0.16 \\ \hline
    \end{tabular}
	\caption{Relevant intrinsic coefficients of the integrated SFMS and RGS relations and of the spatially-resolved SFMS relation, as well as those derived by a linear regression procedure to each of them. The $\sigma$ value is the deviation about the fitting. $^{\textasteriskcentered}$Units: [log(M$_{\odot}$ yr$^{-1}$)]. $^{\dag}$Units: [log(M$_{\odot}$ yr$^{-1}$ Kpc$^{-2}$)]}
	\label{table1}
 \end{center}
\end{table*}

The uncertainties of the derived quantities are dominated by the spectrophotometric accuracy of the CALIFA data, which is estimated to be $\sim$6$\%$ \citep{Garcia15}, as well as the details of the procedure followed to derive the $M_{*}$ and the SFRs . In the case of $M_{*}$ it has been determined that the uncertainties are well constrained within a typical error of $\sim$0.15 dex, \citep[see e.g. the discussion in][]{Sanchez13}. The typical size of the uncertainties for our data are shown in Figure \ref{fig:sfrVSmas_int}, which are consistent with what has been previously found by \citet{CidFernandes14} and \citet{Catalan2015}, and were taken into account when computing the linear fitting.

\subsection{Spatially-Resolved SFMS relation}

\begin{figure}[H]
  \centering
    \includegraphics[width=0.38\textwidth,angle=90]{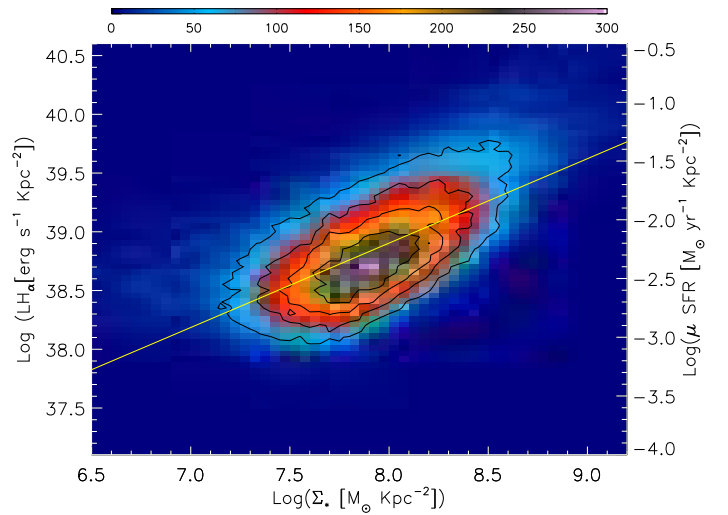}
  \caption{Spatially-Resolved SFMS relation for the CALIFA sample, that holds for scales in the range of 0.5-1.5 kpc. Colors in the plot represent the amount of data points presented in this study. The outermost contour holds within itself the 80$\%$ of the total data presented in the plot. Further contours holds 60$\%$, 40$\%$  and 20$\%$ of the total amount of data in the plot. Yellow line represents the linear fitting to the spatially-resolved SFMS relation using the 80$\%$ of the data (see Table \ref{table1} for further details of the linear fitting).}
  \label{fig:sfrVSmas}
\end{figure}

Next we explore the spatially-resolved SFMS for the CALIFA sample, by plotting the SFR surface density versus the $M_{*}$ surface density in Figure \ref{fig:sfrVSmas}. The individual spaxels used in constructing the relation have spatial  scales of 0.5-1.5 kpc which are larger than  typical HII region sizes  (hundreds of parsecs) \citep{GonzalezDelgado97}. It was derived using only the spaxels that fulfil the same criteria of the blue points in Figure \ref{fig:sfrVSmas_int}, irrespective of the location of its host galaxy in that figure. Indeed, 11$\%$ of the included spaxels comes from galaxies for which the global ionisation is not dominated by SF (i.e., not blue in Fig. \ref{fig:sfrVSmas_int}). 

A density plot of the local SFMS, based on 90,786 individual spectra, is shown in Figure \ref{fig:sfrVSmas}. Colors account for the density of data points, and the yellow line represents a linear fitting to the correlation using only the 80$\%$ of the data, corresponding to the data contained inside the largest contour in the density plot which corresponds to the linear regime of the data. However in Table \ref{table1} we present the results of the fitting using both, 100$\%$ and 80$\%$ of the data. The fitting was performed using variable mass bins which individually contain 1$\%$ of the total amount of data and shows that the spatially-resolved SFMS relation holds in general, as a linear relation. Uncertainties in this plot are not shown, but were taken into account in the computation of the linear fitting and are reflected in the errors of the slope and zero point of the correlation.

The spatially-resolved SFMS proves to be a tight correlation, as the standard deviation ($\sigma$) is quite small $\sigma$ = 0.23 dex, comparable to the value obtained for the integrated relation, in this work and previous independent studies (see for example: D07, S14, S15).

\section{Discussion} \label{Disc}
		
In Section 4.1 we explored the tight correlation between the SFR (inferred from H$\alpha$ luminosity) and the total $M_{*}$ for star-forming (SFMS) and retired (RGS) galaxies, in logarithmic scales. In the case of the integrated SFMS the slope and zero point (See Table \ref{table1}) are in good agreement with the previously reported values. Even if the range of variation of reported slopes and zero points in the literature may be as large as $\sim 0.4 \textendash1$ for the slope and $\sim -(4\textendash10)$ for the zero points for local galaxies (S14), many recent works tend to constrain these values to smaller ranges, such as $0.71\textendash0.77$ and $-(6.78\textendash7.64)$ \citep[RP15]{Zahid12,Elbaz07}. Even more, published dispersions of this relation ($\sim0.2-0.35$ dex), are in good agreement with the dispersion found with our data (0.20 dex). 

In section 4.2 we presented the spatially-resolved SFMS relation. The observed correlation holds on kpc scales, and is consistent with the slope of 0.66 $\pm0.18$ reported by \citet{Sanchez13} for HII regions. Comparing the integrated  and the spatially-resolved relations, our results (summarised in Table \ref{table1}) indicate that not only the slopes, but also the intrinsic scatter, are roughly of the same order in both relations.



 We explored the local relation in different $M_{*}$ bins, finding that it has no dependence with this parameter. We fitted linear correlations to each mass bin relation, and used a 2D Kolmogorv-Smirnov test to verify that the spatially-resolved SFMS present the same distribution irrespective of the mass of the host galaxy, however we will explore this in more detail in further works. We also explored the possible effects in the local SFMS determination, relaxing our criteria to select the SF regions by allowing the regions laying below the KL and whose EW(H$\alpha$) are $\textless$ 3 to be considered also as SF regions. From this test we found that for both relations, using the 100$\%$ and 80$\%$ of the data the slope vary just $\sim$0.04 dex, i.e. it has a little impact in our results.

From a physical point of view, we know that SF is a local process, and therefore it is not unreasonable that the scaling relations governing it are local as well. The fact that the SFMS relation remains as a tight linear correlation on kpc scales suggests that the conversion of gas into stars is mainly driven by local rather than global processes.

A local SFMS could only be derived from the global one if the spatial distribution of both the $\mu$SFR and $\Sigma_{*}$ followed similar patterns in all galaxies. Given the variations in the shape and normalization of these observed profiles \citep[e.g.][Gonz\'{a}lez Delgado et al., in preparation]{GonzalezDelgado15}, it seems rather fine-tuned that a mechanism acting on the scale of the whole galaxy affects both surface densities exactly in the way required to yield a universal relation on resolved scales without a local process being involved.

To recover the global relation from the local one we just need to integrate the resolved quantities across the whole area of the galaxy. In terms of the stellar mass surface density $\Sigma_h$ at the half-light radius $R_h$, the total stellar mass should roughly scale as $ M_* \propto \Sigma_h R_h^2 $. 
If the star formation surface density fulfils $\mu_{SFR} \propto \Sigma_*^\gamma$ at every point (i.e. a local SFMS with logarithmic slope $\gamma$), one obtains the integrated relation $\psi$ $\propto \Sigma_h^\gamma R_h^2 = \Sigma_h^{\gamma-1} M_* $.
Since $ \Sigma_h \propto M_*^\beta $ with $\beta \sim 0.5$ \citep[e.g.][]{Kauffmann2003, GonzalezDelgado15}, the logarithmic slope of the integrated SFMS would be
\begin{equation}
 \alpha = 1 + (\gamma-1)\beta \sim 1 - 0.3 \times 0.5 = 0.85
 \label{eq_alpha}
\end{equation}
if it was a consequence of the local relation.
If equation~(\ref{eq_alpha}) holds, $ \alpha - \gamma = (1-\beta)(1-\gamma) > 0 $ implies that the integrated SFMS should be slightly steeper than the local relation, in rough agreement with the results reported in Table~1.

It must be noted, though, that this prediction is only valid if $ M_* \propto \Sigma_h R_h^2 $, which is merely a first-order approximation. Moreover, the measured value of the logarithmic slope $\alpha$ is rather sensitive to the sample selection criteria (inclination, EW threshold, etc.), and we have to keep in mind that the galaxies used in the integrated relation, even if they were selected for being dominated by the SF activity, may contain local zones that are not star-forming.

		
\section {Conclusions} \label{Conc}

We report the spatially-resolved SFMS relation using IFS data from the CALIFA survey, that holds for kpc scales. Our sample consist of galaxies of mixed morphological types, and masses that extend to three orders of magnitude. $M_{*}$ have been derived from stellar population fits to optical spectra and SFRs have been inferred from the extinction-corrected intensity of the H$\alpha$ emission line

For an integrated SFR vs. $M_{*}$ plot we identified two main sequences, one accounts for the RGS and the other for the SFMS itself, for this last one we report a slope of 0.81 $\pm 0.02$ and a dispersion of 0.20 dex. For the star-forming areas in each galaxy, irrespectively of their integrated properties, we find a correlation between the $\mu$SFR and the $\Sigma_{*}$ that is as tight as the integrated one, and that seems to be the fundamental relation from which the global one is derived. For the local SFMS we found a slope of 0.72 $\pm 0.04$ and a dispersion of  0.23 dex.

In future articles we will explore the possible dependance of this relation with other properties of the galaxies, like morphology, color, environment, etc., as well as the derivation of the spatially-resolved RGS

{\small \textit{Acknowledgements:} The referee, E. P\'erez and R. Cid Fernandes for their comments. Financial support: MCD and SFS: DGAPA-UNAM funding; CONACyT-180125 and PAPIIT IA-100815 projects. ZS: EU Marie Curie Career Integration Grant "SteMaGE" PCIG12-GA-2012-326466. YA: RyC-2011-09461 and AYA2013-47742-C4-3-P projects from the Spanish MINECO and the SELGIFS programme, funded by the EU (FP7-PEOPLE-2013-IRSES-612701). CJW: Marie Curie Career Integration Grant 303912. RMGD: AyA2014-57490-P and JA P12-FQM2828 grants. JFB: AYA2013-48226-C3-1-P from the Spanish MINECO grant. LG: Millennium Science Initiative through grant IC120009, and by CONICYT through FONDECYT grant 3140566.}

\clearpage

\end{document}